\documentclass[sigconf,screen]{acmart}
\usepackage{url}
\usepackage{graphicx}
\usepackage{hyperref}
\usepackage[utf8]{inputenc}
\usepackage{array}
\usepackage{enumitem}
\usepackage{minted}
\usepackage{subcaption}
\usepackage{algorithm}
\usepackage{algpseudocode}
\usepackage{listings,lipsum}
\usepackage[english]{babel}
\usepackage{blindtext}
\usepackage{listings}
\usepackage[skip=1pt,font=small]{caption}
\AtBeginDocument{%
  }

\setcopyright{acmlicensed}
\copyrightyear{2026}
\acmYear{2026}
\acmDOI{XXXXXXX.XXXXXXX}
\acmISBN{978-1-4503-XXXX-X/18/06}

\acmBooktitle{arXiv preprint}
\acmConference[arXiv]{arXiv}{2026}{}

\renewcommand\footnotetextcopyrightpermission[1]{}

\lstdefinestyle{sharpc}{language=[Sharp]C, frame=lr, rulecolor=\color{blue!80!black}}

\begin{document}
    \title{TestMap: Evidence Infrastructure for Foundation-Model-Assisted Test Generation}
    \author{Hunter Leary}
    \email{hunterl22@vt.edu}
    \orcid{0009-0008-1056-8106}
    \affiliation{%
      \institution{Department of Computer Science, Virginia Tech}
      \city{Blacksburg}
      \state{Virginia}
      \country{USA}
    }
    \author{Luke Hanuska}
    \email{lhanuska@vt.edu}
    \affiliation{%
      \institution{Department of Computer Science, Virginia Tech}
      \city{Blacksburg}
      \state{Virginia}
      \country{USA}
    }
    \author{Chris Brown}
    \email{dcbrown@vt.edu}
    \orcid{0000-0002-6036-4733}
    \affiliation{%
      \institution{Department of Computer Science, Virginia Tech}
      \city{Blacksburg}
      \state{Virginia}
      \country{USA}
    }

    \renewcommand{\shortauthors}{Leary et al.}

\begin{abstract}
    Foundation models (FMs) can generate plausible unit tests, but determining whether those tests are correct, useful, maintainable, and worth integrating remains difficult. Generated tests must be mapped to the code they target, inserted into real projects, built, executed, measured against the baseline suite, repaired when necessary, and compared across models and generation strategies. This validation process is often fragmented across build systems, test runners, coverage tools, mutation tools, static analyzers, and experiment scripts. The problem is especially important because generated tests are both code artifacts and validation artifacts: they must themselves be validated before they can be trusted as evidence about the system under test.
    
    This paper presents \textsc{TestMap}, an open-source infrastructure prototype that automates evidence-backed foundation-model-assisted test generation for C\#/.NET repositories. TestMap supports repository analysis, source-test mapping, baseline execution, code metric collection, test smell detection, coverage measurement, mutation testing, model-guided test generation, validation, repair, and repository-specific experiment tracking. Rather than reporting only final passing tests, TestMap records the lifecycle of each generated candidate, including failed, repaired, low-impact, and evidence-positive outcomes. These intermediate outcomes can reveal model limitations, missing context, repair cost, toolchain inefficiencies, or possible faults in the system under test. Using TestMap as a design case, we describe the architecture and evidence model needed to make generated tests observable, repeatable, and comparable across repositories, models, prompts, and generation strategies. We conclude with lessons learned and open challenges, including oracle and assertion quality, metric attribution, test maintainability, flakiness, execution cost, and developer acceptance.
\end{abstract}



\maketitle

\section{Introduction}
\label{sec:introduction}
Foundation models (FMs) are increasingly used to produce software artifacts such as code, tests, and documentation. Stack Overflow's 2025 Developer Survey reports that ``84\% of respondents are using or planning to use AI tools in their development process,'' with ``51\% of professional developers'' using AI tools daily. The same survey reports AI use across common development tasks, including documentation, testing code, and writing code, while also finding that 45\% of respondents distrust AI tool outputs~\cite{stackoverflow_2025_2025}. Google's 2025 DORA report similarly finds that ``90\% of technology professionals report using AI at work,'' yet 30\% report having ``little to no trust in the code generated by AI''~\cite{google_dora_2025}. JetBrains reports a similar pattern: ``85\% [of] developers regularly use AI for coding and development,'' while quality of generated code remains a leading concern~\cite{jetbrains_artificial_2025}. Together, these reports suggest that AI-assisted software development is becoming standard practice, but also introduces an active validation challenge: \textit{FM-generated artifacts are becoming common in development workflows before developers fully trust them}. For generated tests, this validation challenge creates a practical infrastructure problem: researchers and developers need repeatable ways to collect evidence about whether generated tests build, pass, improve testing signals, expose failures, and justify further review.

This validation challenge is especially important for software testing. Tests are not only software artifacts that must compile, execute, and be maintained; they are also evidence artifacts used to support confidence in the behavior of the system under test. When FMs generate tests, especially in workflows where they may also generate implementation code, software teams face a recursive validation problem: generated tests must themselves be validated before they can be trusted to validate the system under test. Previous work shows that generated code can be incorrect and that tests are needed to verify generated code~\cite{liu_is_2023}; work also shows that generated tests may fail to capture code implementation bugs~\cite{mathews_when_2026}.

This validation problem builds on long-standing challenges in software testing. Software testing is essential for improving software quality and reliability, but writing and maintaining tests remains costly. While unit tests help verify the behavior of the system, prior work shows that manual test creation is time-consuming and often avoided by developers~\cite{latoza_maintaining_2006,meyer_today_2021,daka_survey_2014,schafer_empirical_2024}. To reduce the manual test-creation burden, automated test generation tools, like EvoSuite and Randoop, have sought to generate tests with minimal human intervention~\cite{fraser_evosuite_2011,pacheco_feedback-directed_2007}. These tools have proven advantages in terms of increasing code coverage and mutation scores; however, prior work also reports challenges in maintainability, readability, and effectiveness on complex codebases~\cite{shamshiri_how_2018,wang_automatic_2021}. These results highlight a particular theme within automated test generation: producing tests is only one part of the problem; determining whether the generated tests are useful, maintainable, and worth integrating is equally, if not more, important.

FMs renew this challenge by enabling test generation from a variety of sources, including source code, project artifacts, requirements, issue reports, and natural language prompts~\cite{schafer_empirical_2024,rao_cat-lm_2023,nashid_issue2test_2025,wei_requirements_2024,feldt_towards_2023}. Recent work suggests that LLM-based test generation can perform competitively with traditional approaches~\cite{bhatia_unit_2024}, and developer surveys indicate practitioner interest in using AI tools for testing code~\cite{google_dora_2025,stackoverflow_2025_2025}. However, the validation problem remains; a generated test may compile and pass but this does not prove correctness---the test oracle problem~\cite{barr_oracle_2015}. A generated test may provide additional code coverage without improving fault detection. A generated test may also improve mutation score while suffering from test smells. Recent work suggests that LLMs can generate test oracles that are comparable to programmer-written oracles in some settings~\cite{molina_test_2025}. As Alshahwan et al. argue, the growing availability of automated testing tools requires rigorous evaluation to determine whether generated tests are accurate and relevant in practice~\cite{alshahwan_software_2023}.

Signals relevant to this evaluation already exist in software engineering practice. For example, build outcomes, execution results, coverage reports~\cite{hemmati_how_2015}, mutation testing results~\cite{petrovic_does_2021}, static code metrics, and test smells~\cite{spadini_relation_2018} can all provide evidence about \textit{candidate tests}---or test cases automatically generated by FMs to verify program behavior that developers must review and integrate into their codebase. However, these signals are often fragmented across tools and typically reported as aggregate outcomes rather than being connected to the candidate. This fragmentation creates an infrastructure gap: FM-assisted test generation requires systems that can collect, relate, and preserve evidence across the lifecycle of a generated test candidate. Such infrastructure should support candidate-level review, such as whether a generated test appears worth accepting or investigating, and strategy-level analysis, such as which model, prompt, or repair workflow works best for a particular codebase.

We present \textsc{TestMap}, an open-source infrastructure prototype for \textit{evidence-backed FM-assisted test generation and evaluation}. TestMap automates the process of generating, validating, measuring, repairing, and recording generated test candidates in real C\#/.NET repositories. It treats generated tests as candidates rather than final answers and records the lifecycle of each candidate. TestMap supports repository analysis, source-test mapping, baseline test execution, static code metric collection, test smell detection, coverage measurement, mutation testing, context-aware model-guided test generation, validation, repair, and repository-specific experiments for comparing models, prompts, and generation approaches.

TestMap is currently implemented in C\# and targets .NET projects, using this as a concrete setting for studying evidence-backed test generation. However, the underlying approach is not specific to C\#: generated tests in any language must be mapped, validated, measured, repaired, and compared before they can be trusted as test contributions. By mapping generated test candidates to static, dynamic, failure, repair, and generation evidence, TestMap makes generated tests observable, repeatable, and comparable. TestMap is designed to run with minimal repository-specific configuration, allowing developers and researchers to study projects of interest rather than relying only on a fixed set of benchmark repositories.

This paper makes the following contributions:
\begin{enumerate}[topsep=0pt]
    \item We frame FM-assisted test generation as a recursive validation problem: generated tests must be validated before they can be trusted as validation artifacts.
    \item We present \textsc{TestMap}, an open-source evidence infrastructure \item We present \textsc{TestMap}, an open-source evidence infrastructure prototype that automates repository analysis, generated-test validation, evidence collection, repair tracking, and experiment logging for FM-generated test candidates.
    \item We describe how TestMap preserves the full candidate lifecycle, including failed, repaired, and passing tests, so intermediate outcomes can support developer review and empirical analyses.
    \item We detail repository-specific experiments that help researchers and developers compare models, prompts, generation strategies, and repair workflows on a concrete codebase.
    \item We discuss limitations and open challenges for the future of software testing in the FM era, including assertion quality, flakiness, metric attribution, costs, and developer acceptance.
    \label{list:contributions}
\end{enumerate}

\section{Background}
\label{sec:background}

\subsection{Automated Test Generation}
\label{subsec:auto-test-gen}
Automated test generation is an established research area~\cite{garousi_when_2016}. Tools such as Randoop and EvoSuite generate unit tests for Java programs using feedback-directed random testing and search-based software testing, respectively~\cite{pacheco_feedback-directed_2007,fraser_evosuite_2011}. Randoop constructs method-call sequences and uses execution feedback to discard illegal or redundant tests, while EvoSuite uses genetic algorithms to optimize generated suites toward testing objectives such as maximizing coverage. These systems demonstrate that automated techniques can produce executable tests and improve measurable testing signals. However, prior work also shows persistent adoption challenges: generated tests can be difficult for developers to read, understand, and maintain~\cite{daka_generating_2017,shamshiri_how_2018}. These limitations motivate a recurring theme: generated tests should not be evaluated only by whether they execute or improve aggregate metrics, but also by whether they are understandable, useful, and suitable for integration into a project.

\subsection{Foundation-Model-Assisted Test Generation}
\label{subsec:fm-test}
Recent work has explored machine learning and foundation models for test generation~\cite{tufano_unit_2020,tufano_generating_2022,alshahwan_automated_2024,rehan_harnessing_2025,schafer_empirical_2024}. Compared with traditional random or search-based testing techniques, FM-assisted techniques can incorporate diverse forms of context, including focal methods, documentation, existing test results, and natural language instructions~\cite{rao_cat-lm_2023,nashid_issue2test_2025,wei_requirements_2024,feldt_towards_2023}. For example, Rao et al. introduce CAT-LM, a model designed for test generation, and compare it with existing code generation models as well as traditional automated testing tools~\cite{rao_cat-lm_2023}. Their approach builds on previous research that extracts focal context by mapping tests to methods they exercise, using this information when training the model~\cite{tufano_unit_2020}. RUG is a recent FM-assisted test generation tool that uses contextual information built from static analysis to generate unit tests for the Rust programming language~\cite{cheng_rug_2025}. However, practitioners note challenges leveraging LLMs for software testing~\cite{santana_software_2025}.

\subsection{Evaluating Generated Tests}
\label{subsec:eval-gen-test}

Evaluating generated tests is difficult because common testing signals capture different and incomplete aspects of test quality. A generated test that compiles and passes is executable in the target project, but passing alone does not imply that the test checks meaningful behavior. This reflects the broader test oracle problem: tests require a mechanism for determining whether the observed behavior of the system under test is correct, and automating this judgment remains a central challenge in software testing~\cite{barr_oracle_2015}.

Coverage is a commonly used signal for evaluating generated tests. While useful, coverage is incomplete because it shows which code was exercised, not whether the exercised behavior was validated by meaningful assertions~\cite{hemmati_how_2015}. Mutation testing provides a stronger signal by measuring whether tests detect injected behavioral changes~\cite{petrovic_does_2021}, but mutation testing is computationally expensive and does not fully establish semantic correctness~\cite{pizzoleto_systematic_2019}. Test smells and code metrics provide additional evidence about maintainability and readability~\cite{paul_xnose_2024}. Prior work also shows that production code can be more defect-prone when tested by smelly tests~\cite{spadini_relation_2018}.

Failed generated tests are also evidence. A failing candidate may reflect an invalid oracle, missing context, or a model limitation, but it may also reveal ambiguous behavior, a possible implementation defect, or a weakness in the generation pipeline. Recent work suggests that design choices in FM-assisted test generation tools can cause bug-revealing tests to be discarded or can lead generated tests to validate faulty implementation behavior~\cite{mathews_design_2024}. These signals are often fragmented across tools, limiting the ability to connect generated tests with the evidence needed to validate them.

\subsection{C\# and the .NET Testing Ecosystem}
\label{subsec:csharp-test}
Although much automated test generation research has focused on Java and Python, C\# is widely used in practice and remains important in enterprise software development~\cite{github_staff_octoverse_2025,tiobe,crossover}. C\# also provides a practical setting for studying evidence-backed test generation because the .NET ecosystem includes mature infrastructure for static analysis,\footnote{https://github.com/dotnet/roslyn} test execution, coverage collection,\footnote{https://learn.microsoft.com/en-us/dotnet/core/testing/unit-testing-code-coverage} and mutation testing.\footnote{https://stryker-mutator.io/docs/stryker-net/introduction/} Earlier work on C\# test generation includes Pex, later integrated into Visual Studio as IntelliTest, which uses dynamic symbolic execution to generate tests~\cite{tillmann_pex_2008}. More recent work has explored transformer-based unit test generation for game development scenarios involving C\#~\cite{paduraru_unit_2024}, and xNose provides test smell detection for C\# test suites~\cite{paul_xnose_2024}.

\subsection{Evidence-Based Software Engineering}
\label{subsec:evidence}
Evidence is critical for enhancing the reliability of decisions and practices across disciplines~\cite{barends_evidence-based_2018,pennington_evidence_1986}. In particular, \textit{contextual evidence} gathers information from local data sources~\cite{bornstein_putting_2017,CDC-context}. Research outlines the need for evidence-based software engineering practices~\cite{dyba_evidence-based_2005}. For instance, Dyba compares software development to the medical domain, motivating the need for enhanced organizational and technical infrastructure to promote evidence-based practices~\cite{dyba_evidence-based_2005}. Prior work finds developers often rely on experiences to guide development practices rather than evidence~\cite{passos_analyzing_2011}, and suggests contextual evidence is often unavailable~\cite{freund_contextualizing_2015,salvaneschi_context-oriented_2012}. Further, studies show lack of contexts inhibits LLMs from effectively supporting software development tasks, reducing test case quality~\cite{hai_impacts_2025}. Meanwhile, techniques to improve contextual evidence, such as prompt engineering~\cite{altmayer_pizzorno_coverup_2025,ouedraogo_prompt_2026} and project contexts~\cite{anderson_improving_2014,tufano_unit_2020}, can enhance the automatically generated test cases. We extend these efforts by introducing evidence infrastructure for FM-assisted test case generation, leveraging contextual evidence to assist developers in assessing the performance of FM-generated test cases.

\paragraph{Summary}
Taken together, prior work suggests FM-assisted test generation requires more than producing executable tests. Traditional test generation shows that measurable improvements do not necessarily imply maintainable or useful tests. FM-assisted approaches expand the context available for generation, but preserve challenges around oracle quality, validation, and trust. Existing testing tools produce useful signals, but those signals are often fragmented across tools and disconnected from individual generated candidates. These observations motivate the need for evidence-based FM-assisted test generation infrastructure that can connect generated test candidates to repository context, validation outcomes, failure histories, and generation strategy-level comparisons.
\section{TestMap}
\label{sec:method}

\subsection{Overview}
\label{subsec:testmap-overview}
\begin{figure*}[ht]
    \centering
    \includegraphics[width=1\linewidth]{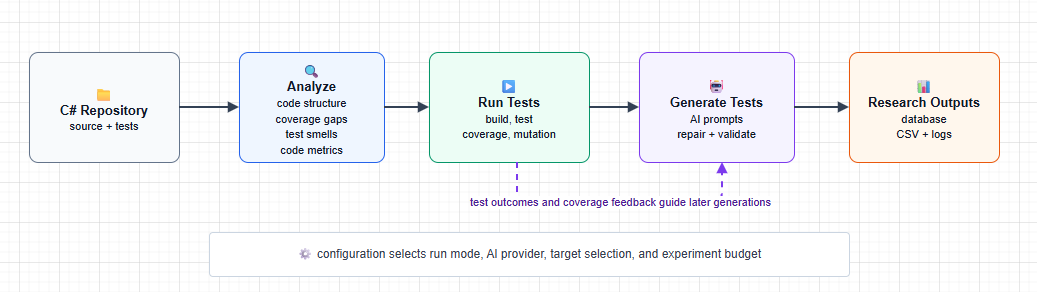}
    \caption{Overview of TestMap Generation Pipeline}
    \label{fig:testmap-outline}
\end{figure*}

\textsc{TestMap} is an open-source evidence infrastructure prototype for foundation-model-assisted test generation. The current implementation targets C\#/.NET repositories, but the underlying workflow (see Figure~\ref{fig:testmap-outline}) can generalize to other programming languages 

At a high level, TestMap ingests a repository, analyzes its source and test projects, records baseline testing evidence, selects generation targets, constructs model prompts, generates candidate tests, validates those candidates through build and execution, collects coverage and mutation evidence, detects code metrics and test smells, and persists candidate histories for analysis. Each generated test is associated with evidence about its target, generation context, execution outcome, testing contribution, quality indicators, repair history, and final status. This evidence allows developers and researchers to inspect not only whether a generated test passes, but also what it covers, whether it improves the existing suite, whether it detects behavioral changes, whether it required repair, and whether it appears maintainable.

TestMap is designed to support both candidate-level and strategy-level decisions. At the candidate level, the collected evidence can help developers decide whether a generated test should be accepted, edited, investigated, or rejected, although the current prototype does not yet study developer decisions directly. At the strategy level, TestMap's experiment workflow enables repository-specific comparisons of models, prompts, context settings, generation strategies, pass@k budgets, and repair loops. In this way, TestMap complements standardized benchmarks by providing evidence about how foundation-model-assisted test generation behaves in a concrete codebase with its actual build system, dependencies, existing tests, and quality signals.

\subsection{Evidence Model}
\label{subsec:evid-model}
TestMap organizes evidence around the lifecycle of a generated test candidate. A candidate test begins as model output, but it is not treated as a test contribution until it has been validated, measured, and associated with evidence about its behavior in the target repository. We use the term \textit{evidence-backed} to mean that each generated candidate is linked to structured signals about its target, generation context, execution behavior, testing contribution, quality indicators, failure history, repair history, and pipeline outcome. These signals do not prove that a generated test is correct; rather, they make the candidate inspectable and comparable.

The generated test candidate is the central evidence-bearing object in TestMap. Each candidate is linked to a source-code method target, along with contextual information such as the containing class, uncovered regions, survived mutants, and related tests. It is also associated with the generation context that produced it, including the model, prompt, generation strategy, context mode, attempt number, history setting, and repair budget. After generation, TestMap validates the candidate in the target repository and records evidence from build and execution, coverage, mutation testing, code metrics, test smell detection, failure diagnostics, and repair attempts.

Table~\ref{tab:evidence-model} summarizes the major evidence categories collected by TestMap and the questions they support.

\begin{table*}[ht]
\centering
\caption{Evidence categories collected by TestMap.}
\label{tab:evidence-model}
\begin{tabular}{p{0.22\linewidth}p{0.36\linewidth}p{0.34\linewidth}}
\toprule
\textbf{Evidence Type} & \textbf{Examples} & \textbf{Question Supported} \\
\midrule
Repository evidence & Commit, project path, target framework, test framework, baseline state & Where and under what conditions was the candidate evaluated? \\
Target evidence & Source method/class, coverage gap, survived mutant, related tests & What behavior was the candidate intended to exercise? \\
Generation evidence & Model, prompt, strategy, context, attempt number, pass@k setting & How was the candidate produced? \\
Execution evidence & Build result, test outcome, runtime errors, logs & Does the candidate run in the target project? \\
Testing-impact evidence & Coverage delta, killed mutants, survived mutants, baseline comparison & Does the candidate add observable value beyond the baseline suite? \\
Quality evidence & Code metrics, test smells, complexity indicators & Does the candidate appear maintainable and reviewable? \\
Failure evidence & Compiler errors, assertion failures, exceptions, failed attempts & What did unsuccessful candidates reveal? \\
Repair evidence & Repair attempts, diagnostics, final status, changed failures & What did it take to make the candidate valid? \\
Strategy evidence & Aggregated outcomes by model, prompt, strategy, pass@k, repair loop & Which approach appears effective for this repository? \\
\bottomrule
\end{tabular}
\end{table*}

\paragraph{Failed candidates as evidence.}
TestMap preserves failed and intermediate candidates rather than reporting only final passing tests. A failed candidate may indicate an invalid generated test, but it may also expose missing context, an incorrect oracle, ambiguous behavior, a possible implementation defect, or inefficiency in the generation workflow~\cite{mathews_design_2024}. For example, repeated compile failures may suggest that the prompt lacks project-specific test conventions, while repeated assertion failures may indicate weak oracle inference or a mismatch between inferred and implemented behavior~\cite{chen_chatunitest_2024}. Environment failures, timeout failures, and unsuccessful repair attempts can also reveal limitations in the execution harness or repair strategy. Treating failed candidates as evidence allows TestMap to support analysis of both generated-test quality and toolchain effectiveness.

\paragraph{Pipeline outcomes.}
TestMap assigns each candidate a pipeline outcome that summarizes its validation state. These outcomes are derived from collected evidence and should not be interpreted as developer acceptance. A candidate may be marked as \textit{ValidationFailed} when it does not compile, execute, or pass validation; \textit{Validated} when it compiles, executes, and passes but has not yet been classified by impact; \textit{ValidatedLowImpact} when it validates but produces no observed improvement over the baseline according to configured evidence signals; or \textit{ValidatedEvidencePositive} when it validates and improves at least one configured signal, such as coverage or mutation results.

\subsection{Pipeline Stages}
\label{subsec:pipeline-stages}

\paragraph{Repository Ingestion}
TestMap begins by cloning the target repository and recording provenance information, including the current branch and commit hash. After cloning, TestMap searches for .NET solution files, beginning with solution formats such as \texttt{.sln} and \texttt{.slnx}. TestMap uses Roslyn APIs to open the solution programmatically and discover the C\# projects contained in it. For each project, TestMap records project files, dependencies, target frameworks, build targets, project references, and included C\# source files. This repository and project metadata is persisted in SQLite and used by later analysis, execution, generation, and attribution stages.

\paragraph{Static Analysis}
Following repository ingestion, TestMap performs static analysis on the source and test files discovered for each project. Static analysis records dependency information, such as using directives, and structural information, including namespaces, classes, structs, records, constructors, methods, and properties. TestMap also collects code metrics, including lines of code and cyclomatic complexity, using a custom library based on Roslyn's compiler APIs. Additionally, TestMap applies a customized version of xNose to detect test smells in C\# test code~\cite{paul_xnose_2024}. TestMap records relationships among program elements, including inheritance, method invocations, and source-test relationships where they can be resolved. These results are persisted in SQLite and provide the structural context used for target selection, prompt construction, and later evidence attribution.

\paragraph{Baseline Execution}
After static analysis, TestMap attempts to execute the repository's existing test suite to establish a baseline before generated tests are introduced. TestMap first restores dependencies and builds the project; if either step fails, later generation results cannot be meaningfully attributed to generated tests. Execution is performed in a configured environment, typically a custom Docker image based on the official .NET SDK image with additional testing tools installed. TestMap supports both Linux and Windows container images when project dependencies require different execution environments. During baseline execution, TestMap runs the existing tests, collects test results in TRX format, and collects coverage reports in Cobertura XML format. When feasible, TestMap also runs Stryker.NET to collect mutation testing results.

\paragraph{Result Attribution}
After baseline execution, TestMap attributes execution artifacts back to the static program model. Coverage reports are mapped to source files, methods, lines, and branches where possible. TRX results are mapped back to discovered test methods. Mutation reports are mapped to source locations, mutation types, and mutant outcomes, including killed and surviving mutants. This attribution step is important because later generation stages rely on evidence such as uncovered code, related tests, and survived mutants to construct context and select targets.

\paragraph{Target Selection}
TestMap selects source elements as generation targets using configurable criteria. In the current workflow, targets are typically source methods, although the same process can be applied to other source elements such as classes, constructors, or properties. A target represents the code element for which TestMap will attempt to generate or improve test coverage. Target selection may be guided by baseline evidence, such as existing related tests, uncovered lines or branches, surviving mutants, code metrics, or other indicators of testing weakness. TestMap then retrieves related evidence from the database, including the target source code, containing type, related tests, coverage information, mutation results, code metrics, test smells, dependencies, and invocation relationships. This evidence is used to construct the generation context.

\paragraph{Test Generation}
TestMap uses a decomposed generation pipeline, inspired by RUG, rather than asking the model to produce a complete test in a single prompt~\cite{cheng_rug_2025}. For each selected source target, TestMap first constructs an evidence package from the repository database. This package includes the target source code, related tests, test framework information, coverage or mutation gaps, test-class context, helper/setup members, dependencies, and available quality metadata. The evidence package is then converted into a test generation request.

The generation pipeline may optionally build a context graph and resolve that graph into prompt-level construction hints. These steps are deterministic preprocessing steps: they organize available source, test, fixture, dependency, and metric evidence before model generation begins. The model-facing generation sequence then proceeds through a series of planning steps. Under the \textsc{NoHistory} context mode, each step is prompted independently using the selected evidence and relevant previous artifacts supplied explicitly in the prompt. Under the \textsc{ChainedHistory} context mode, TestMap also carries a conversation transcript forward so that later steps can condition on earlier model decisions.

\begin{table}[t]
\centering
\caption{TestMap generation steps.}
\label{tab:generation-steps}
\begin{tabular}{p{0.28\linewidth}p{0.62\linewidth}}
\toprule
\textbf{Step} & \textbf{Purpose} \\
    \midrule
    Evidence package & Select repository evidence relevant to the target, including source code, related tests, metrics, framework conventions, dependencies, and coverage or mutation gaps. \\
    Context graph & Optionally organize available evidence into a graph of target members, dependencies, fixture/setup helpers, constructors, factories, and candidate construction hints. \\
    Context resolution & Optionally convert graph nodes into concrete prompt hints, such as reusable setup snippets or preferred object-construction paths. \\
    Scenario planning & Ask the model to identify a focused behavior, edge case, branch, or missing scenario that the generated test should exercise. \\
    Method name generation & Ask the model to produce a test method name consistent with the scenario and the target project's test framework conventions. \\
    Arrangement planning & Ask the model to plan required objects, mocks, fixtures, helper reuse, namespaces, and setup strategy. \\
    Input planning & Ask the model to choose concrete inputs and preconditions needed to exercise the selected scenario. \\
    Action planning & Ask the model to plan the invocation under test and identify how the result should be bound. \\
    Assertion planning & Ask the model to plan expected behavior and framework-specific assertions. \\
    Final test generation & Ask the model to synthesize the complete C\# test method from the scenario, method name, plans, context, examples, and project conventions. \\
    Repair generation & When repair is enabled, ask the model to revise a failed candidate using compiler, build, execution, assertion, coverage, or structured diagnostic feedback. \\
    \bottomrule
\end{tabular}
\end{table}

For experiment runs, TestMap persists the generation steps associated with each attempt, including the step type, status, prompt, response, response format, structured response when available, prompt version, validation status, token counts, duration, and error or skip metadata. Some steps may be disabled or represented by deterministic fallback artifacts rather than model responses. Repair attempts are stored as separate generation attempts linked to the original candidate.

\paragraph{Pre-Execution Candidate Validation}
Before executing the modified test suite, TestMap validates whether the generated candidate can be represented, inserted, compiled, and analyzed. This stage includes:
\begin{enumerate}
    \item \textbf{Syntax extraction:} TestMap attempts to extract a valid C\# syntax tree from the model output.
    \item \textbf{Insertion validation:} TestMap inserts the candidate into the selected test project and verifies that the file and test structure are placed correctly.
    \item \textbf{Compiler diagnostics:} TestMap invokes Roslyn compilation checks to identify syntax, type, reference, or framework errors before execution.
\end{enumerate}
Candidates that pass pre-execution validation are statically analyzed using the same analysis pipeline as existing tests. TestMap records their structural elements, code metrics, test smells, and relationships to source elements.

\paragraph{Post-Execution Candidate Validation}
After pre-execution validation, TestMap evaluates the candidate by executing the modified test suite. This stage follows the baseline execution pattern but targets the modified test project containing the generated candidate. TestMap records whether the project restores, builds, executes, and passes. It also collects TRX test results,\footnote{The default format for .NET test results, \url{https://www.nuget.org/packages/Microsoft.Testing.Extensions.TrxReport}} coverage reports, and, when feasible, mutation results for the modified suite. These results are compared against the baseline to compute candidate-level evidence, such as coverage deltas, newly covered code, killed or surviving mutants, and execution failures. Failed executions are preserved with diagnostics because they may indicate invalid candidates, incorrect oracles, environment issues, or possible defects in the system under test.

\paragraph{Evidence Persistence}
All artifacts produced during the pipeline are persisted in the repository database. This includes repository provenance, static program structure, baseline execution results, candidate selection evidence, prompts and model outputs, generated test code, validation diagnostics, repair attempts, execution outcomes, coverage and mutation deltas, code metrics, test smells, and pipeline outcomes. Persisting this evidence allows TestMap to preserve the full lifecycle of each candidate, including failed, repaired, low-impact, and evidence-positive outcomes. It also enables repository-specific experiments by aggregating candidate outcomes across models, prompts, context modes, generation strategies, pass@k budgets, and repair loops.

\subsection{Repository-Specific Experiments}
TestMap treats test generation as a repository-specific process rather than a model-only benchmark. The same prompt or model can behave differently across repositories because projects differ in test frameworks, fixture conventions, dependency structure, build requirements, coverage gaps, mutation profile, and existing test style. For this reason, TestMap records experiments at the level of a concrete repository revision and evaluates generated candidates in the target repository itself.

Each experiment is associated with repository provenance, including the repository identifier, branch, commit hash, discovered solution or project files, target framework, and baseline execution results. Candidate selection is performed against this repository state. In the current workflow, TestMap selects source methods using configurable evidence such as baseline coverage, surviving mutants, related tests, static metrics, and risk or metric-opportunity scores. Each selected method becomes a target for test candidate generation.

An experiment expands these targets across a generation matrix. Matrix dimensions may include the model provider, generation approach, metrics path, context mode, generation budget, temperature, and enabled generation steps. For example, one experiment may compare naive prompting against metrics-driven prompting, coverage-only evidence against mutation-aware evidence, or independent-step prompting against chained-history prompting. Step-ablation experiments disable selected planning stages to estimate their contribution to candidate outcomes.

For each matrix item and target method, TestMap executes a generation budget. A pass@1 configuration produces one candidate. A pass@k configuration produces multiple independent candidates. A repair configuration first generates an initial candidate and then performs bounded repair attempts using validation diagnostics from earlier attempts. Repair prompts may include compiler errors, structured build diagnostics, test failures, assertion failures, runtime logs, coverage feedback, and the prior generation transcript when chained history is enabled.

Generated candidates are inserted into the target repository, validated, executed, measured, and then rolled back between attempts when the experiment uses an isolated rollback workspace. TestMap persists every attempt rather than only the best passing candidate. For each attempt it records the generated code, generated method name, prompts, model responses, step metadata, token counts, validation diagnostics, execution results, coverage and mutation deltas, repair linkage, and final pipeline outcome. This makes failed, low-impact, repaired, and evidence-positive candidates available for later analysis.

The output of a repository-specific experiment is both a database record and an analysis file. The database preserves detailed nested artifacts, while the result file provides one row per generation or repair attempt with stable identifiers back to the persisted evidence. Rows include the repository revision, target method, baseline state, generation configuration, model/provider, attempt number, repair metadata, validation status, testing-impact signals, quality indicators, failure category, and pipeline outcome. Aggregating these rows allows TestMap to compare generation strategies within a repository and to study how repository characteristics affect generated-test outcomes. This experimentation capability allows users to make evidence-based decisions on FM-generated candidates and tools in testing workflows.

\section{Discussion}
\subsection{TestMap as a Complement to Benchmarks}
Standardized benchmarks are valuable because they provide controlled settings for comparing models and test generation techniques. However, benchmark performance does not necessarily answer the repository-specific question developers often face: which model, prompt, or generation strategy produces useful tests for this codebase? Benchmarks may become outdated, may be affected by data contamination, and may abstract away project-specific factors such as build systems, dependencies, testing frameworks, existing test suites, and coding conventions.

TestMap is intended to complement, rather than replace, benchmarks. Its repository-specific experiment workflow allows developers and researchers to compare models, providers, prompts, context settings, generation strategies, and repair workflows on a concrete project. This provides evidence about how FM-assisted test generation behaves under the actual constraints of the target repository rather than only under curated benchmark conditions.

\subsection{Failed Candidates as Evidence}

Failed generated tests should not always be treated as discarded model outputs. A failing candidate may reflect an invalid oracle, missing context, or a model limitation, but it may also expose ambiguous behavior, an implementation defect, or a weakness in the generation pipeline. This is especially important for source-driven test generation. If a tool keeps only generated tests that pass, it may bias the process toward tests that confirm the current implementation rather than tests that challenge it. Recent work similarly suggests that design choices in FM-assisted test generation tools can cause bug-revealing tests to be discarded or can lead generated tests to validate faulty implementation behavior~\cite{mathews_design_2024}.

TestMap preserves failed and repaired candidates as first-class evidence. These outcomes can help distinguish candidate errors from model limitations, toolchain issues, ambiguous behavior, and possible defects in the system under test. They also reveal the cost of generation: a passing candidate produced after multiple failures and repairs provides different evidence than a candidate that validates on the first attempt.

\subsection{Developer Decision Support}

TestMap is designed to produce evidence that can support developer review, although the current prototype does not directly study developer decisions. For each generated candidate, TestMap can surface information about the target source element, generation context, build and execution outcomes, coverage and mutation deltas, code metrics, test smells, failure diagnostics, and repair history. This evidence can help a developer decide whether a generated test appears worth accepting, editing, investigating, or rejecting.

The same evidence may also help developers reason about the existing test suite. Baseline coverage, mutation results, code metrics, and test smell reports provide a view of current testing strengths and weaknesses before generated tests are introduced. Studying how developers interpret this evidence, and whether metrics improve or miscalibrate trust in generated tests, remains future work.

\subsection{Implications for Empirical Test Generation Research}

As Alshahwan et al. argue, the growing availability of automated testing tools creates a need for rigorous and unbiased evaluation of generated tests~\cite{alshahwan_software_2023}. TestMap contributes to this direction by preserving candidate-level evidence rather than reporting only aggregate outcomes. This enables researchers to study not only whether a model can produce passing tests, but also which candidates failed, which required repair, which improved coverage or mutation results, which introduced quality issues, and which strategies produced useful outcomes for particular repositories.

This evidence is also important as common testing metrics are incomplete (see Section~\ref{subsec:eval-gen-test}). Coverage and mutation score are useful signals, but do not fully establish oracle correctness, assertion quality, maintainability, or alignment with intended behavior. A central concern for future FM-assisted test generation research is whether generated tests merely conform to the existing implementation or help reveal what the implementation should do. TestMap provides infrastructure for studying this question by preserving generated candidates, failure histories, repair trajectories, and repository-specific evidence.

\section{Limitations and Future Work}
\label{sec:limit_future}

\subsection{Limitations}
\label{subsec:limitations}

\paragraph{Language and ecosystem scope.}
The current TestMap implementation targets C\#/.NET projects. This provides a practical setting because the .NET ecosystem includes mature infrastructure for static analysis, test execution, coverage collection, and mutation testing. However, parts of the implementation are language- and ecosystem-specific, including Roslyn-based analysis, TRX parsing, Cobertura mapping, and Stryker.NET integration.

\paragraph{Repository and environment support.}
TestMap is designed to run on real repositories, but repository execution remains difficult. Projects may require specific SDK versions, operating systems, external services, credentials, databases, or build steps. Some failures may therefore reflect environment configuration rather than generated-test quality.

\paragraph{Assertion and oracle quality.}
TestMap collects execution outcomes, coverage, mutation results, code metrics, and test smells, but these signals do not fully establish assertion quality or oracle correctness. A generated test may compile, pass, and improve coverage while encoding an incorrect or weak oracle. Mutation testing can reveal some weak tests, but it is not a complete measure of semantic correctness.

\paragraph{Flakiness and stability.}
The current pipeline validates generated tests through build and execution, but a candidate that passes once may still be flaky. Generated tests may depend on timing, test order, shared state, randomness, external resources, or environment-specific behavior.

\paragraph{Developer acceptance.}
TestMap provides evidence intended to support developer review, but the current prototype does not directly evaluate how developers interpret or act on that evidence. As a result, TestMap can support developer decision-making, but it does not yet measure whether its evidence improves review quality, trust calibration, or acceptance decisions.

\subsection{Future Work}
\label{subsec:future-work}

\paragraph{Generalizing beyond C\#/.NET}
Future work should investigate how the evidence model and pipeline generalize to other languages and ecosystems, including Java, Python, Rust, and Go. This will require replacing or extending language-specific components such as Roslyn-based analysis, .NET test execution, and Stryker.NET mutation testing with ecosystem-appropriate alternatives.

\paragraph{Improving repository and environment support.}
Future work should improve environment inference, container selection, dependency handling, and failure classification. Better environment support would help distinguish failures caused by generated tests from failures caused by project setup, missing dependencies, operating-system assumptions, or external services.

\paragraph{Assessing assertion and oracle quality.}
Future work should incorporate assertion-specific analyses, specification alignment, and developer review to better evaluate whether generated tests validate intended behavior. This is especially important because passing tests may still encode weak or incorrect oracles.

\paragraph{Detecting flakiness.}
Future work should add repeated-run validation, randomized test ordering, and cross-environment execution to better characterize the stability of generated tests. These additions would help distinguish robust generated tests from candidates that pass only under narrow execution conditions.

\paragraph{Studying developer review and trust.}
Future studies should examine whether coverage, mutation, smell, failure, and repair evidence improve developer decisions or instead create overconfidence in weak generated tests. This would help determine how evidence should be presented to developers during generated-test review.

\paragraph{Conducting empirical studies.}
TestMap enables repository-specific experiments comparing models, prompts, context modes, generation strategies, pass@k budgets, and repair workflows. Future work will use this infrastructure to conduct large-scale evaluations across open-source repositories and to study whether candidate histories can support model selection, ranking, repair policies, or fine-tuning.

\section{Conclusion}
\label{sec:conclusion}

Foundation-model-assisted test generation does not eliminate the difficulty of software testing; it shifts the challenge toward validating, measuring, repairing, and interpreting generated test candidates. This shift is especially important because generated tests are both code artifacts and validation artifacts: they must be trusted before they can be used to support trust in the system under test.

This paper presented \textsc{TestMap}, an open-source infrastructure prototype for evidence-backed FM-assisted test generation. TestMap records the lifecycle of generated test candidates by connecting repository context, source-test mappings, baseline execution, static metrics, test smells, coverage, mutation testing, validation outcomes, failure histories, repair attempts, and generation metadata. By preserving failed, repaired, and passing candidates, TestMap supports candidate-level review and repository-specific comparison of models, prompts, and generation strategies.

We envision future FM-assisted testing tools that not only generate tests, but also provide the evidence needed to decide whether those tests are useful, maintainable, trustworthy, and worth integrating. TestMap is a step toward that evidence infrastructure, complementing standardized benchmarks with repository-specific analysis of generated test behavior.

\section{Data Availability}
\label{sec:availability}
Our tool, \textbf{TestMap}, and related documentation is available on GitHub at \href{https://github.com/consulthunter/TestMap}{consulthunter/TestMap}. It is also archived on Zenodo at \href{https://doi.org/10.5281/zenodo.14262974}{https://doi.org/10.5281/zenodo.14262974}.

\section*{Acknowledgments}
The author used generative AI tools, OpenAI ChatGPT, for brainstorming, outlining, and editing portions of this manuscript. The author also used OpenAI Codex for implementation assistance, including code refactoring and development support for the TestMap prototype. All generated text, code, claims, citations, and implementation changes were reviewed, edited, and validated by the author, who takes full responsibility for the content of this work.

\bibliographystyle{acm}
\bibliography{references,dcbrown}

\end{document}